\documentclass[10pt]{iopart}
\pdfoutput=1
\usepackage{iopams}
\usepackage{graphicx}
\usepackage{dsfont}
\usepackage{bm}
\usepackage{mathrsfs}
\usepackage[T1]{fontenc}
\usepackage[usenames]{color}
\usepackage[colorlinks=true,breaklinks=true,allcolors=blue]{hyperref}
\usepackage{cite}

\newcommand\one{{\mathds{1}}}

\begin{document}
\title{$N$-Scaling of Timescales in Long-Range $N$-Body Quantum Systems}
\author{Michael Kastner}
\address{National Institute for Theoretical Physics (NITheP), Stellenbosch 7600, South Africa}
\address{Institute of Theoretical Physics, Department of Physics, University of Stellenbosch, Stellenbosch 7600, South Africa}
\ead{\mailto{kastner@sun.ac.za}}

\begin{abstract}
Long-range interacting many-body systems exhibit a number of peculiar and intriguing properties. One of those is the scaling of relaxation times with the number $N$ of particles in a system. In this paper I give a survey of results on long-range quantum spin models that illustrate this scaling behaviour, and provide indications for its common occurrence by making use of Lieb-Robinson bounds. I argue that these findings may help in understanding the extraordinarily short equilibration timescales predicted by typicality techniques.
\end{abstract}

\date{\today}

\section{Introduction}
\label{s:introduction}

Understanding and predicting the relevant timescales on which a certain dynamical phenomenon takes is place is in general a hard task. While analytical perturbative expansions or numerical simulation techniques can deal to some extend with dynamics on the shorter timescales, the slow timescales are usually harder to deal with. The slowest timescale of a many-body system is the one on which relaxation to thermal equilibrium occurs, a topic that has seen renewed interest in recent years. On the applied side, the interest is to a large extend due to experimental realisations of many-body quantum systems that are well isolated from their environment (see \cite{GeorgescuAshhabNori14} and references therein). On the theoretical side, quantum information and typicality methods have lead to substantial progress in the field, cumulating in rigorous proofs of equilibration (in some suitable sense) in generic many-body quantum systems \cite{Reimann08,Linden_etal09,Goldstein_etal10,Short11,ReimannKastner12,ShortFarrelly12,GogolinEisert16}. While these results establish equilibration in principle, i.e.\ after a sufficiently long time, they provide no, or only limited, information on the relevant timescale. For the foundations of statistical mechanics, i.e.\ the question of why Boltzmann-Gibbs distributions are ubiquitous in nature, the timescale of equilibration is essential: If equilibration happened only on an unphysically long time, the above mentioned proofs would not justify from first principles why Boltzmann-Gibbs distributions are so frequently encountered. In recent years the question of equilibration timescales has been addressed in a number of papers \cite{ShortFarrelly12,GoldsteinHaraTasaki13,Malabarba_etal14,GoldsteinHaraTasaki15,Reimann16}, but the results leave some important questions open.

One possible way of gaining insight into the problem is by studying classes of systems that are known to exhibit peculiar behaviour of equilibration timescales, as understanding the origin of these peculiarities may help identifying the crucial ingredients and mechanisms at play. One such class are long-range interacting systems, i.e.\ systems with pair interactions that decay like $1/r^\alpha$ with the inter-particle distance $r$, as their equilibration times $\tau$ have been observed to scale with the system size $N$ like $\tau\propto N^{q}$ \cite{CamDauxRuf09,Kastner11,Kastner12,vdWorm_etal13,KastnerVdWorm15}. Depending on the details of the system, the exponent $q$ can be positive or negative, implying a diverging or vanishing equilibration time in the large-$N$ limit. These extreme equilibration times, which become either extremely short or extremely long for large but finite systems, make long-range interacting models a fertile testbed for the study of equilibration timescales.

This paper is a written account of a talk given at the StatPhys26 conference in Lyon in 2016. Section \ref{s:equilibration} contains an example of a result that proves equilibration in an isolated quantum system in a probabilistic sense, and I briefly discuss the role of the equilibration timescale. In section \ref{s:Ising} I show anecdotal evidence of the $N$-scaling of equilibration timescales in the long-range quantum Ising model. Section \ref{s:LiebRobinson} gives a brief review of Lieb-Robinson bounds, which are general and mathematically rigorous upper bounds on the speed at which physical effects can propagate in space and time. These bounds are not directly applicable to equilibration timescales, but are concerned with dynamical phenomena of a different kind. Interestingly though, they confirm for general quantum spin models the $N$-scaling found for equilibration times in the long-range Ising model. This result makes it seem plausible that the observed scaling is a general property or long-range systems, not only applying to equilibration times, but also to dynamical phenomena of other sorts. Implications of the reviewed results for the general understanding of equilibration timescales are discussed in section \ref{s:discussion}.

\section{Equilibration of isolated macroscopic quantum systems}
\label{s:equilibration}

It is a general expectation, and a postulate of thermodynamics, that two macroscopic bodies that are brought in contact with each other will thermalise, i.e.\ they will evolve towards a state with a common temperature and, after a sufficiently long time, will be described by a Gibbs state. Elementary as this statement may seem, it is not easily reconciled with the underlying microscopic theory. For concreteness, consider a quantum system consisting of $N$ constituents, with a total Hilbert space 
\begin{equation}
\mathscr{H}=\bigotimes_{i=1}^N \mathscr{H}_i,
\end{equation}
where the dimension of the local Hilbert spaces $\mathscr{H}_i$ is assumed to be finite. Given a Hamiltonian $H$, eigenstates are denoted by $|n\rangle$ and energy eigenvalues by $E_n$, such that $H|n\rangle=E_n|n\rangle$. The time evolution of the expectation value of an arbitrary observable $O$ on $\mathscr{H}$ can be written as
\begin{equation}\label{e:Ot}
\langle O\rangle(t)=\Tr\left[O\rho(t)\right]=\sum_{m,n}\langle m|\rho|n\rangle\exp\left[\rmi(E_m-E_n)t\right]\langle n|O|m\rangle,
\end{equation}
where $\rho$ is the initial density operator of the system. The right-hand side of \eref{e:Ot} is a quasiperiodic function, which implies that $\langle O\rangle$ will return arbitrarily closely to its initial value $\Tr[O\rho]$ infinitely many times, and at arbitrarily late times \cite{BocchieriLoinger57} (similar to the Poincar\'{e} recurrence theorem in classical mechanics \cite{Poincare1890}). As a consequence, $\rho(t)$ will certainly not converge to any equilibrium density operator in the long-time limit. What can happen though is that equilibration occurs in a weaker, probabilistic sense. Loosely speaking, probabilistic equilibration implies that the expectation values \eref{e:Ot} of physically relevant observables, measured with finite resolution, will be indistinguishable from the equilibrium value of the observable for the overwhelming majorities of times. While being weaker in a mathematical sense, such a notion of equilibration is perfectly satisfactory on physical grounds.

To illustrate how such a result looks like, I state and discuss the equilibration result of Ref.~\cite{ReimannKastner12}. The object of study is
\begin{equation}\label{e:tdeltaO}
t_{\delta O} = \frac{1}{T}\int_0^T\rmd t\,\Theta\left(\left|\Tr[O\rho(t)]-\Tr[O\omega]\right|-\delta O\right),
\end{equation}
which is the relative frequency at which, during a time interval between 0 and $T$, the expectation value of the observable $O$ differs from its equilibrium value $\Tr(O\omega)$ by more than a given measurement accuracy $\delta O$. Here $\Theta$ denotes the Heaviside step function. The only reasonable candidate for the ``equilibrium density operator'' $\omega$ is the infinite-time average (or {\em diagonal ensemble} \cite{Polkovnikov11}) of the initial density operator $\rho$,
\begin{equation}
\omega=\sum_n|n\rangle\langle n|\rho|n\rangle\langle n|.
\end{equation}
This equilibrium density operator $\omega$ will in general differ from the microcanonical density operator, or any other of the familiar Gibbs equilibrium ensembles. {\em Ther\-mal\-i\-sa\-tion} in isolated systens, i.e.\ equilibration towards the microcanonical ensemble, can occur only for restricted sets of observables (local and/or macroscopic), a fact that becomes evident by considering the nonlocal observable $|n\rangle\langle n|$, which is invariant under the time evolution and hence will not thermalise. Without additional restrictions on the set of permitted observables, equilibration towards the diagonal ensemble is the most one can hope to prove in isolated quantum systems.

The main result of \cite{ReimannKastner12}, valid after sufficiently long times $T$, but otherwise derived under very mild assumptions, is an upper bound on the relative frequency \eref{e:tdeltaO},
\begin{equation}\label{e:t_bound}
t_{\delta O} \leqslant 6g\left(\frac{\Delta O}{\delta O}\right)^2\max_np_n.
\end{equation}
By finding conditions under which the right-hand side of this inequality becomes very small, measurable deviations from equilibrium are shown to be exceedingly rare, and hence equilibrium values will be measured with extremely high probability.
 
The quantities on the right-hand side of \eref{e:t_bound} can be estimated on physical grounds. $\Delta O$ (which is essentially given by the operator norm $\|O\|$) is the range of possible measurement outcomes of observable $O$, which for any realistic measurement apparatus is finite. The ratio $\Delta O/\delta O$ quantifies the relative precision of the measurement. For a measurement precision of 50 relevant digits one has $\Delta O/\delta O=10^{50}$, and it seems unlikely that a precision as high as this will be reached in the foreseeable future. $p_n=\langle n|\rho|n\rangle$ is the initial population of the $n$th energy level. Generically, due to the exponential growth of the Hilbert space dimension with the particle number $N$, energy levels in a macroscopic quantum system will be extremely numerous, and hence extremely dense on the energy axis. As a consequence, even when most carefully preparing an initial state, the number of populated energy eigenstates will be of the order of $10^{\mathcal{O}(N)}$. Initial populations of energy eigenstates are then expected to be not larger than $\max_np_n=10^{-\mathcal{O}(N)}$, which, for a macroscopic particle number $N$, gives a mind-bogglingly small number. The final remaining ingredient of the bound in \eref{e:t_bound} is
\begin{equation}
g=\max_{(m,n)}\left|\left\{(m',n')\,|\,E_m-E_n=E_{m'}-E_{n'}\right\}\right|,
\end{equation}
where $(m,n)$ denotes pairs of different indices $m\neq n$ (and analogously for primed indices). $g$ quantifies how many of the energy gaps $E_m-E_n$ are the same, which in turn quantifies the degeneracies of oscillation frequencies in the expectation value \eref{e:Ot}. For an integrable model with a high degree of symmetry, $g$ can become large, but for a generic nonintegrable system the degeneracy is expected to be of order one. In this latter case and for a macroscopic system one finds that the right-hand side of \eref{e:t_bound} gives a mind-bogglingly small number of the order of $10^{-\mathcal{O}(N)}$. While deviations from equilibrium do occur in principle, they are either so very small that they cannot be measured, or so exceedingly rare that they will not be observed in practice. The rigorous bound \eref{e:t_bound} together with the above considerations proves equilibration of generic isolated macroscopic quantum systems for experimentally realistic observables and initial states. For a more detailed discussion see \cite{ReimannKastner12}.

What remains unspecified in the above result is the timescale of equilibration. The time interval $[0,T]$ has to be ``sufficiently long'', which can be rephrased as ``there exists a finite $T$ sufficiently large such that \eref{e:t_bound} holds'', but not more than that is said. If $T$ turned out to be as long as the age of the universe, the bound \eref{e:t_bound}, while proving equilibration in principle, would not be a valid justification for the fact that  equilibration is frequently observed in everyday situations. Understanding the timescale of equilibration for generic macroscopic systems is therefore an open question of importance for the foundations of statistical mechanics.

\section{Dynamics of the long-range quantum Ising model}
\label{s:Ising}

While general statements about equilibration timescales are certainly desirable, model calculations can be helpful for developing intuition. The long-range quantum Ising model in a longitudinal magnetic field is simple enough to afford exact analytic solutions for the time-evolution of expectation values of local observables, including spin expectation values \cite{Emch66,Kastner11,Kastner12,BachelardKastner13} as well as spin--spin correlation functions \cite{vdWorm_etal13,KastnerVdWorm15}. I will review here the simplest of those results, extract the equilibration timescales, and discuss the scaling of these times with the system size $N$.

The Ising model in a longitudinal magnetic field of strength $h$ with general coupling $J_{ij}$ is defined by the Hamiltonian 
\begin{equation}\label{e:Ising}
H=-\frac{1}{2}\sum_{i\neq j}J_{ij}\sigma_i^z\sigma_j^z-h\sum_i\sigma_i^z,
\end{equation}
where $\sigma_i^z$ denotes the $z$-component of the Pauli spin operator on lattice site $i$. Owing to the generality of the coupling, this Hamiltonian can be used to describe arbitrary lattices in any dimension. The model is simple in that all operators occurring in \eref{e:Ising} mutually commute, and analytical calculations are feasible without too much effort. Analytical solutions take on a particularly simple form when the initial state is chosen diagonal in the $\sigma^x$-eigenbasis \cite{Emch66},
\begin{equation}\label{e:rho0}
\rho=\frac{1}{2^N}\biggl(\one+\sum_{i}\sigma_i^x\biggl(s_i + \sum_{j>i}\sigma_j^x\biggl(s_{ij}
 + \sum_{k>j}\sigma_k^x\biggl(s_{ijk} + \sum_{l>k}\cdots\biggr)\biggr)\biggr)\biggr)
\end{equation}
where the various $s$ are expansion coefficients. Starting from any initial state of this form, the time evolution of the expectation value of an $x$-Pauli operator is given by \cite{Emch66,Kastner11,BachelardKastner13}
\begin{equation}\label{e:exact}
\langle\sigma_i^x\rangle(t)=\Tr(\sigma_i^x\rho)\cos(2ht)\prod_{j\neq i} \cos \left(2J_{ij}t\right).
\end{equation}

\begin{figure}\centering
 \includegraphics[width=0.48\linewidth]{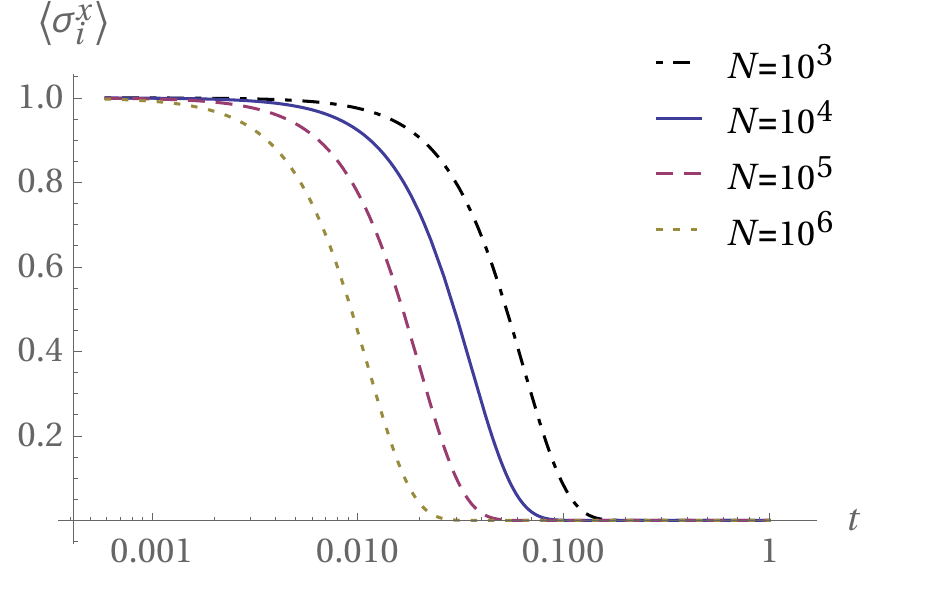}
 \includegraphics[width=0.48\linewidth]{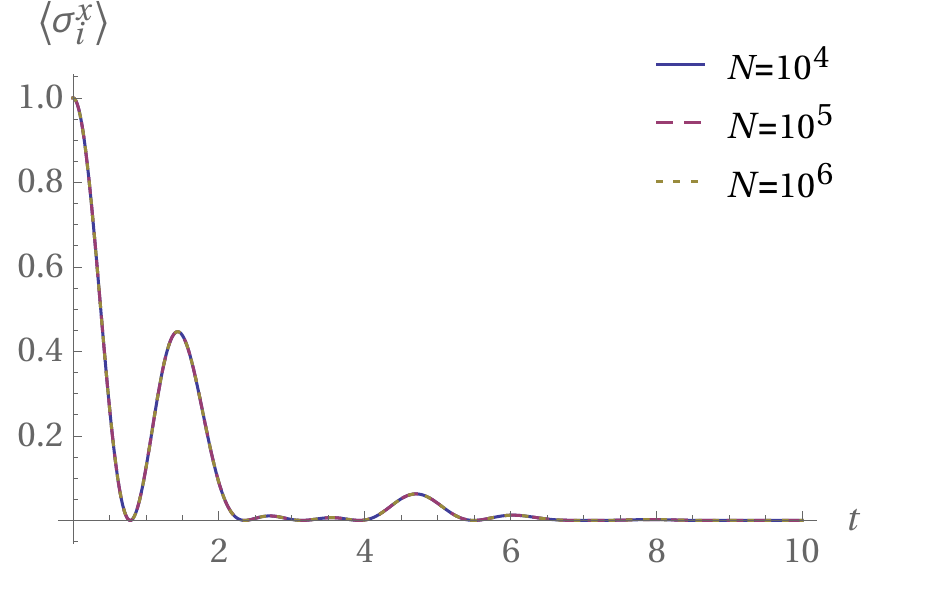}
  \caption{\label{f:sxt}
  Time-evolution of the expectation value $\langle\sigma_i^x\rangle$ for the long-range Ising model \eref{e:Ising} with $h=0$, starting from a fully $x$-polarised product initial state. In both plots $\langle\sigma_i^x\rangle$ relaxes to its zero equilibrium value for all system sizes shown (see legend). Left: For $\alpha=1/4$ relaxation occurs on a timescale that strongly depends on the system size $N$, becoming faster with increasing $N$. Qualitatively similar behaviour is observed for all $0\leqslant\alpha\leqslant1/2$. Right: For $\alpha=2$ relaxation is essentially independent of $N$. Qualitatively similar behaviour is observed for all $\alpha>1/2$.
  }
\end{figure}

As a simple example, consider a one-dimensional lattice and spin--spin couplings that decay like a power law with the distance, $J_{ij}=1/|i-j|^\alpha$, where $|i-j|$ is the distance between lattice sites $i$ and $j$. The exponent $\alpha$ tunes the interaction range, from all-to-all coupling at $\alpha=0$ to nearest-neighbour coupling in the limit $\alpha\to\infty$. Equation \eref{e:exact} is plotted in figure~\ref{f:sxt} for different values of $\alpha$. While $\langle\sigma_i^x\rangle$ relaxes to zero for all values of $\alpha$, the timescale at which this equilibration occurs depends on the system parameters. Figure~\ref{f:sxt} (right), which is for $\alpha=2$, shows an apparent\footnote{The decay is called apparent since, as discussed in section \ref{s:equilibration}, rare large fluctuations away from the zero equilibrium value will occur at later times.} decay to zero, superimposed by oscillations. The plot shows results for different system sizes $N$, ranging over several orders of magnitude, but they all lie on top of each other and cannot be discerned on the scale of the plot. Qualitatively similar behaviour is found for all $\alpha>1/2$. Figure~\ref{f:sxt} (left) is for $\alpha=1/4$, and the apparent decay to zero is again clearly visible, albeit without superimposed oscillations. The main difference compared to $\alpha>1/2$ is the strong dependence of the equilibration timescale on the system size $N$ (note the logarithmic scale of the $t$-axis). The larger $N$, the faster is the approach to equilibrium, and one can read off from the equidistantly (on the log-scale) decaying curves that the equilibration time $\tau$ scales like a power law with the system size, $\tau\propto N^q$. The exponent $q=\alpha-1/2$, valid for all $0\leqslant\alpha<1/2$, can be extracted by means of scaling plots, and the value is also confirmed analytically by an asymptotic evaluation of \eref{e:exact} for large $N$ in the long-time limit \cite{BachelardKastner13}. In agreement with figure~\ref{f:sxt} (left), these negative $q$-values imply that the equilibration timescale becomes faster for larger systems, and goes to zero in the limit $N\to\infty$. Similar calculations can be done for other observables, e.g.\ correlation functions \cite{vdWorm_etal13}, where the relaxation to equilibrium is more complicated and takes place in two steps on two different timescales, but the general conclusions on the $N$-scaling of equilibration times persist.\footnote{It is not uncommon in statistical physics to make the Hamiltonian of a long-range model extensive by introducing an $N$-dependent coupling constant $\propto N^{1-\alpha/D}$ (where $D$ is the lattice dimension) in front of the first sum in \eref{e:Ising}. While this modifies the scaling law of the relaxation times, it does not eliminate their $N$-dependence \cite{BachelardKastner13}.}

In a more general setting, and in particular for entangles initial states, one finds a threshold value $\alpha=1$ (instead of $\alpha=1/2$) at which the equilibration times switch between different types of behaviour \cite{EisertvdWormManmanaKastner13}: For $\alpha>1$ (or, in general, $\alpha$ greater than the lattice dimension) equilibration times are essentially independent of the lattice size $N$. For $\alpha<1$ (or, in general, $\alpha$ smaller than the lattice dimension) equilibration times scale like a power law with $N$ and vanish in the thermodynamic limit.

\section{Lieb-Robinson bounds for long-range quantum systems}
\label{s:LiebRobinson}

For more complicated models, equilibration times are difficult to analyse. A powerful analytical tool is however available for characterising another dynamical phenomenon, namely the spreading of correlations (or excitations or information) in time and space. This tool, now known as Lieb-Robinson bound, was originally proposed in \cite{LiebRobinson72} for unitarily evolving lattice models with finite-range interactions. Generalizations to classical lattice models \cite{Marchioro_etal78,MetivierBachelardKastner14}, open quantum systems \cite{Poulin10,BarthelKliesch12}, general networks \cite{NachtergaeleSims06}, and other settings have been derived subsequently. 

\begin{figure}\centering
 \includegraphics[width=0.7\linewidth]{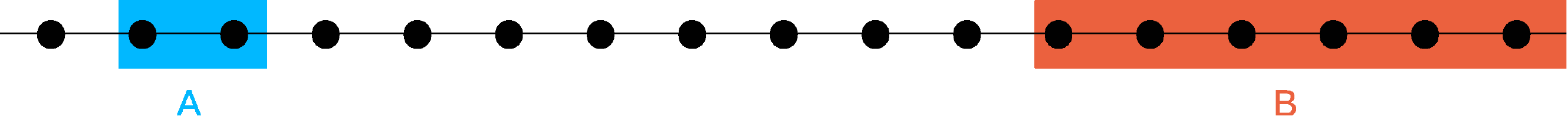}
  \caption{\label{f:lattice}
  Illustration of a one-dimensional lattice $\Lambda$ with two disjoint regions $A$ and $B$.
  }
\end{figure}

\begin{figure}\centering
 \includegraphics[width=0.49\linewidth]{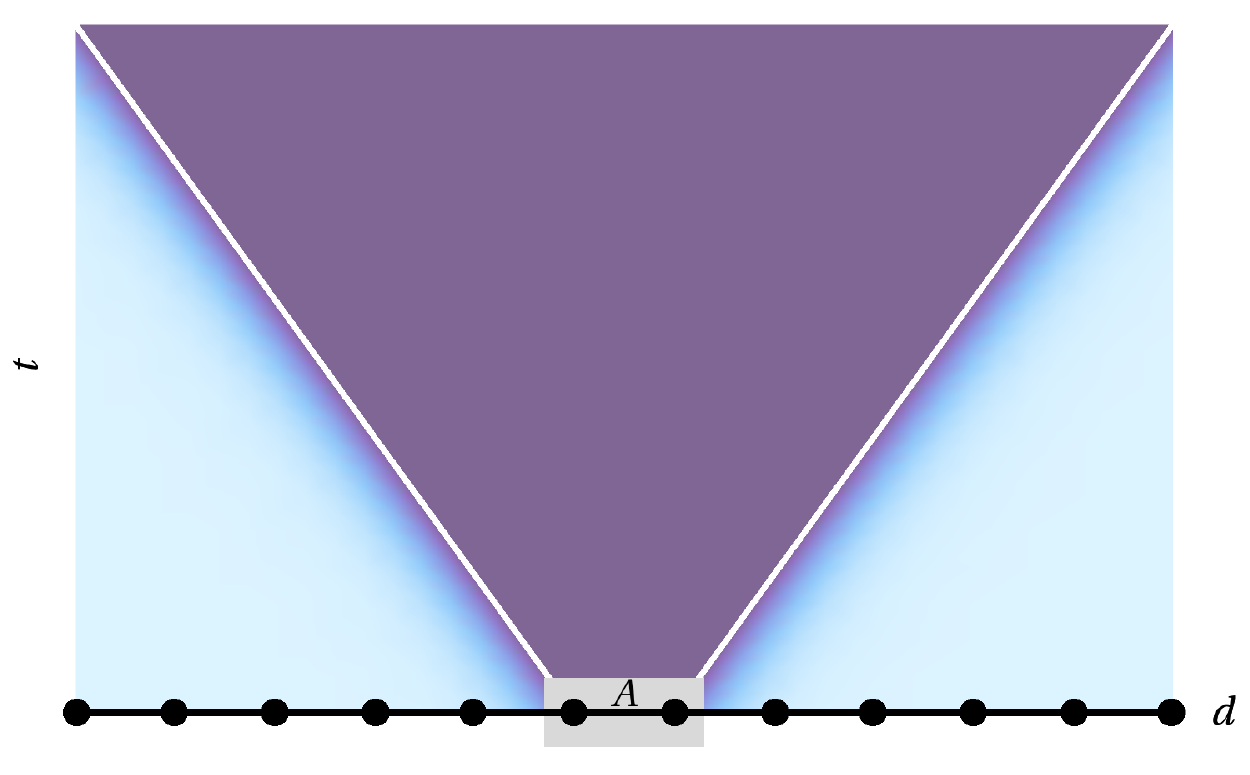}
 \includegraphics[width=0.49\linewidth]{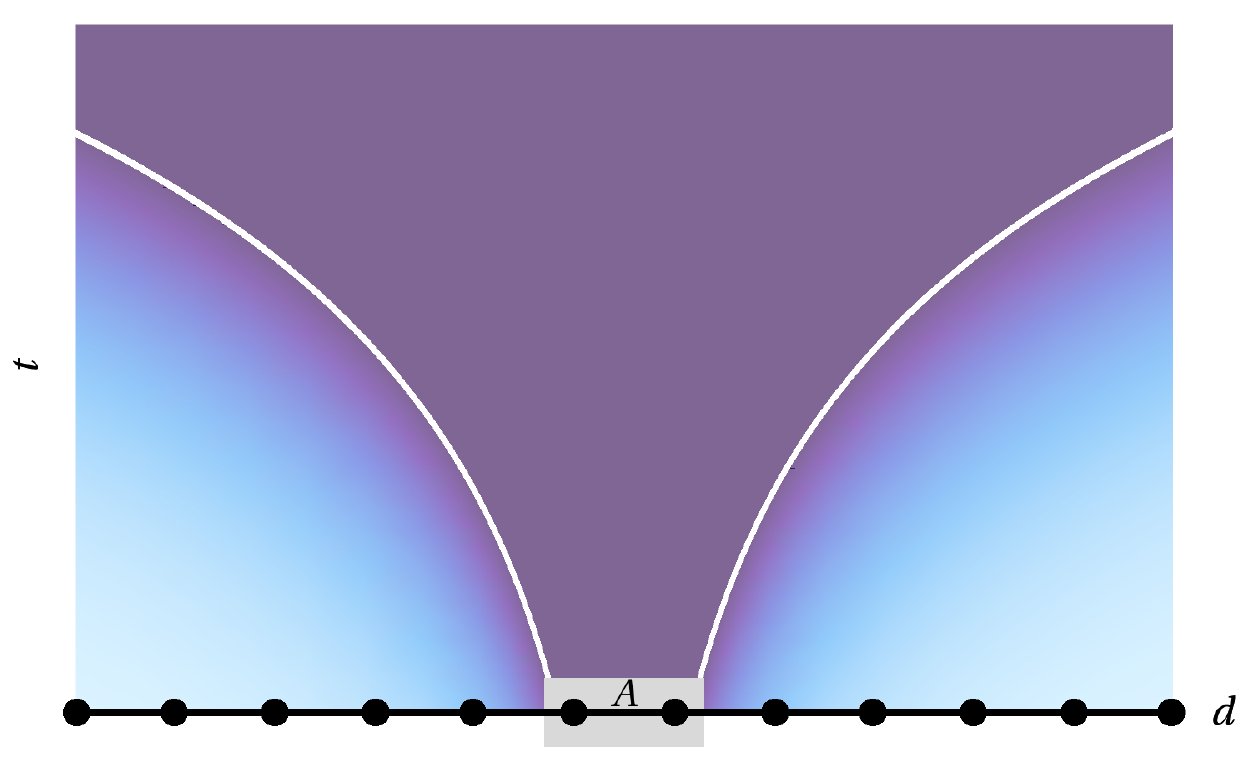}
  \caption{\label{f:cone}
  Illustration of Lieb-Robinson bounds as a function of time $t$ and distance $d$. Darker colours indicate larger values. Left: For finite-range interactions, as in Lieb and Robinson's original paper \cite{LiebRobinson72}, the right-hand side of \eref{e:LRsr} is exponentially small outside a cone-shaped region. Right: For power-law decaying interactions with exponent $\alpha>D$, the right-hand side of \eref{e:LRlr} is small outside a causal region with curved boundaries. 
  }
\end{figure}

A simple setting to give a flavour of Lieb-Robinson bounds is a regular lattice $\Lambda$ with a finite-dimensional Hilbert space $\mathscr{H}_i$ attached to each site $i\in\Lambda$. The tensor product space $\mathscr{H}=\bigotimes_{i\in\Lambda}\mathscr{H}_i$ is then the Hilbert space of the quantum mechanical lattice model, and the model's time evolution is generated by some Hamiltonian $H$ on $\mathscr{H}$. The main objects of study are operators $O_A$ and $O_B$ acting nontrivially only on the (disjoint) subsets $A,B\subset\Lambda$ of the lattice; see figure~\ref{f:lattice} for an illustration. $d(A,B)$ is the distance (in 1-norm) between the regions $A$ and $B$, and $|A|$ and $|B|$ denote the numbers of sites of the regions. In their original work \cite{LiebRobinson72}, Lieb and Robinson derived a bound that, in a less rigorous notation, can be stated as
\begin{equation}\label{e:LRsr}
\left\|\left[O_A(t),O_B(0)\right]\right\|
\leqslant  C \left\| O_A\right\| \left\| O_B\right\| |A| |B| \rme^{(v|t|- d(A,B))/\xi},
\end{equation}
asymptotically for large times $t$ and distances $d$. The object that is bounded is (the operator norm of) the commutator of the operators $O_B$ and $O_A$, the latter time-evolved in the Heisenberg picture, $O_A(t)=\rme^{\rmi Ht}O_A\rme^{\rmi Ht}$. The bound on the right-hand side of \eref{e:LRsr} contains the nonnegative constants $C$, $v$, and $\xi$, which are determined by the lattice structure and the interaction strength. $\|\cdot\|$ denotes the operator norm. The most interesting term in the bound is the exponential, which in the exponent contains time $t$ and distance $d$ in a linear relationship. For any fixed time $t$, the bound is exponentially decaying in $d$ for distances $d>v|t|$. This implies that the commutator on the left-hand side of \eref{e:LRsr} is essentially restricted to a causal region reminiscent of the lightcone in special relativity, with exponentially small corrections outside the cone; see figure~\ref{f:cone} (left) for an illustration. This is a remarkable observation: nonrelativistic quantum mechanics of short-range lattice models has approximately (i.e., except for exponentially small corrections) the same locality structure as a relativistic quantum field theory obeying a finite speed of light. 

A bound on the commutator $\left\|\left[O_A(t),O_B(0)\right]\right\|$ turns out to be a useful tool for deriving bounds on the propagation of information, the building-up of connected correlations, or the creation of entanglement \cite{LiebRobinson72,BravyiHastingsVerstraete06,Kastner15,NachtergaeleOgataSims06}. Less obviously, Lieb-Robsinon bounds like \eref{e:LRsr} can also be used to constrain static properties, e.g.\ the spatial decay of correlations of groundstates of gapped spin models \cite{Hastings04c,HastingsKoma06,NachtergaeleSims06} or of thermal states of fermionic systems \cite{Hastings04b}. 

The first extension of Lieb-Robinson bounds to systems with long-range inter\-actions is due to Hastings and Koma \cite{HastingsKoma06}. Roughly speaking, for a model in $D$ dimensions with pair interactions as in \eref{e:Ising}, the bound
\begin{equation}\label{e:LRlr}
\left\|\left[O_A(t),O_B(0)\right]\right\|
\leqslant  C \left\| O_A\right\| \left\| O_B\right\| |A| |B| \frac{\rme^{v|t|}-1}{(d(A,B)+1)^\alpha}
\end{equation}
holds for couplings $J_{ij}$ that decay sufficiently fast with the distance on the lattice, $J_{ij}\leqslant Cd(i,j)^{-\alpha}$ with $C\geq0$ and $\alpha>D$; see \cite{HastingsKoma06,NachtergaeleOgataSims06} for precise statements and conditions. In the bound \eref{e:LRlr}, time and distance no longer occur in a linear relationship, and as a result the effective causal region is not a cone: Defining the causal region as the part of the $(d,t)$-plane where the right-hand side of \eref{e:LRlr} is smaller than some threshold value, one finds a region with logarithmically curved boundaries; see figure~\ref{f:cone} (right) for an illustration. This curved shape indicates that the concept of a finite group velocity, familiar from condensed matter theory, may break down in the presence of long-range interactions. For a fixed time $t$, the bound \eref{e:LRlr} decays asymptotically like $d^{-\alpha}$ outside the causal region, and hence much slower than in \eref{e:LRsr}. A refined bound due to Foss-Feig \etal \cite{FossFeigGongClarkGorshkov15} for $\alpha>2$ shows that the curved boundaries of the causal region are given by power laws, not logarithms, with an exponent that is a nontrivial function of the long-range exponent $\alpha$.

\begin{figure}\centering
 \includegraphics[width=0.32\linewidth]{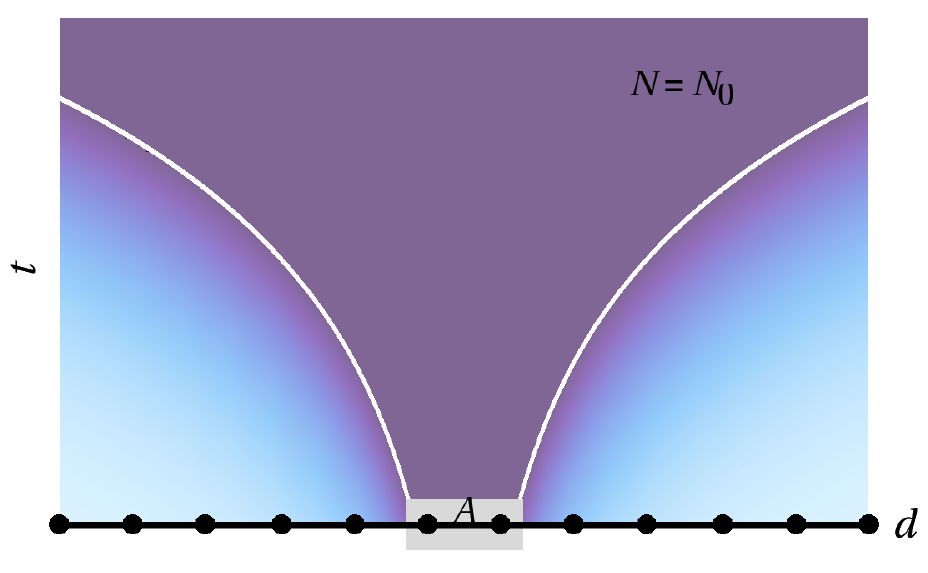}
 \includegraphics[width=0.32\linewidth]{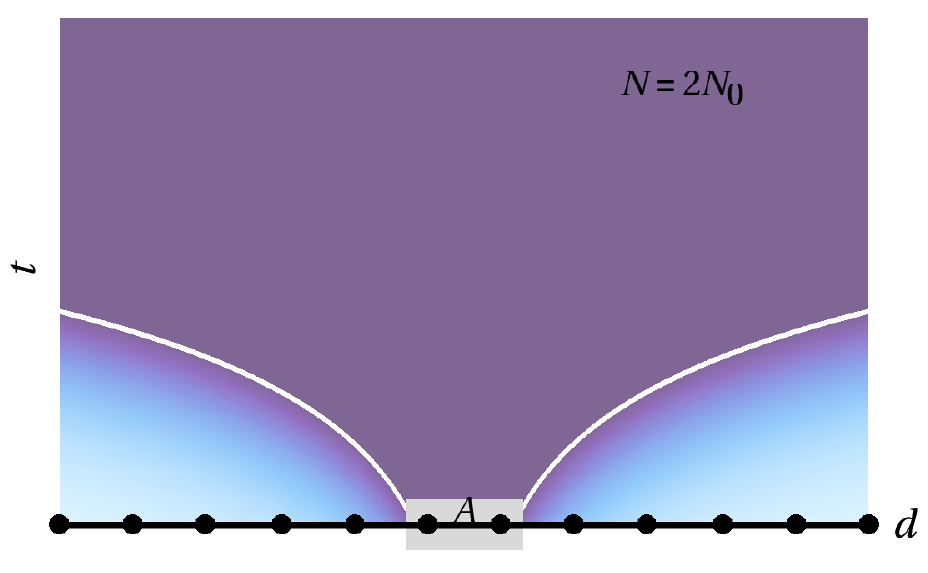}
 \includegraphics[width=0.32\linewidth]{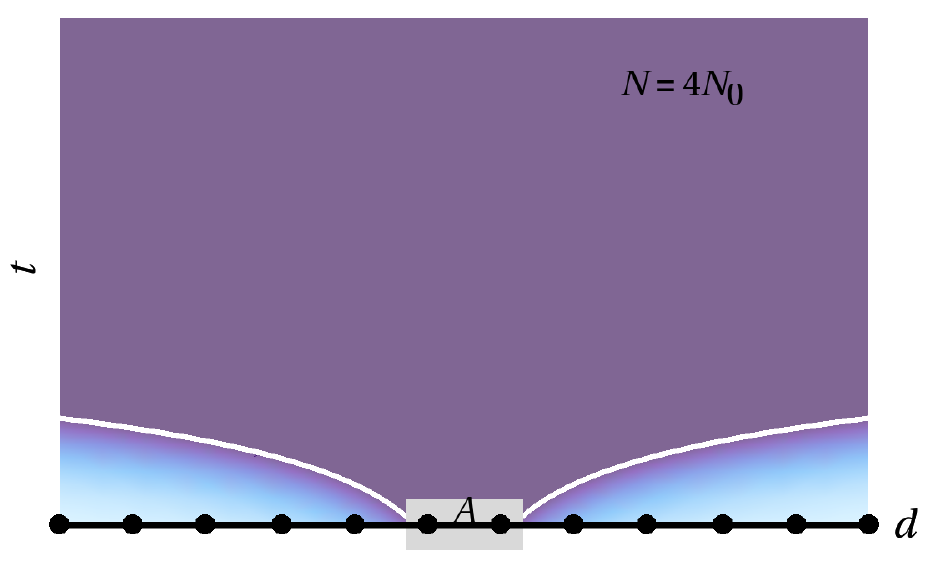}
  \caption{\label{f:coneScaled}
  Illustration of the Lieb-Robinson bound \eref{e:LRlrTau}, valid for $0\leqslant\alpha<1$, for different system sizes $N=N_0$, $2N_0$, and $4N_0$ (from left to right). Darker colours indicate larger values. In equation \eref{e:LRlrTau} the bound, which is independent of $N$, is expressed in rescaled time $\tau$, and this corresponds to an $N$-dependent behaviour in real time $t$.
  }
\end{figure}

Hasting and Koma's bound \eref{e:LRlr} is not valid for exponents $\alpha$ smaller than the lattice dimension $D$. To understand the reason for this restriction it is important to note that, while Lieb-Robinson bounds are valid also for finite systems, they seek to provide estimates valid for long times and large distances. A meaningful bound is therefore expected to remain valid in the thermodynamic limit of an infinitely extended lattice. In section~\ref{s:Ising} it was shown for the long-range Ising model that, for long-range exponents $\alpha<D$, the relaxation time $\tau$ scales with the system size $N$ like $\tau\propto N^q$ with $q<0$. Assuming that a similar kind of $N$-scaling also applies to the dynamics described by the Lieb-Robinson commutator, one immediately arrives at two conclusions: (a) A Lieb-Robinson bound of the type \eref{e:LRlr} with $N$-independent coefficients $C$ and $v$ cannot be valid, because the unlimited speed-up of the dynamics with increasing $N$ would at some point violate any such bound. (b) The problem can be remedied, and a finite bound can be derived, by an appropriate rescaling of time. Such a result for $0\leqslant\alpha<D$ has been obtained by Storch \etal \cite{StorchvandenWormKastner15},
 \begin{equation}\label{e:LRlrTau}
\!\!\!\left\|\left[O_A(\tau N^{\alpha/D-1}),O_B(0)\right]\right\|
\leqslant  C \left\| O_A\right\| \left\| O_B\right\| |A| |B| \frac{\rme^{v|\tau|}-1}{(d(A,B)+1)^\alpha}.
\end{equation}
The right-hand side is a finite bound with $N$-independent constants, valid also in the thermodynamic limit. The bound predicts, similar to the result of Hastings and Koma, a curved causal region as illustrated in figure~\ref{f:cone} (right). Additionally, it also describes the acceleration of the dynamics with increasing system size $N$: The right hand side of \eref{e:LRlrTau} is independent of $N$ when expressed in terms of the rescaled time $\tau$. In real time $t=\tau N^{\alpha/D-1}$, however, time evolution proceeds more rapidly with increasing $N$; see figure~\ref{f:coneScaled} for an illustration. Taking these observations together, the bound \eref{e:LRlrTau} provides estimates for the shape of the causal region, and also for the acceleration of the dynamics with increasing system size.

In section \ref{s:Ising} it was found that for the long-range Ising model with exponents $0\leqslant\alpha<1$ the equilibration time scales like a power law with the system size and vanishes in the limit $N\to\infty$. The result of the present section, albeit not dealing with equilibration timescales, places this observation on a broader basis: In the general, model-independent framework of Lieb-Robinson bounds one likewise finds power law scaling with the system size $N$, precisely in the parameter regime $0\leqslant\alpha<1$ for which also the equilibration times of the long-range Ising model show such behaviour. While this does not constitute a proof, the observations point towards a general phenomenon that occurs in systems with long-range interactions.

\section{General results on the timescales of equilibration}
\label{s:discussion}

Based on the observations made in sections \ref{s:Ising} and \ref{s:LiebRobinson}, it is tempting to speculate that extremely fast dynamics is a common phenomenon in systems with long-range interactions. At this point I would like to close the circle and get back to the topics of equilibration and typicality discussed in section \ref{s:equilibration}. As mentioned in the introduction, typicality techniques have been used to address the question of equilibration timescales of typical systems and/or typical observables \cite{Malabarba_etal14,GoldsteinHaraTasaki15,Reimann16}. The settings discussed in these papers differ from each other, but in all cases it was observed that equilibration (or thermalisation in \cite{GoldsteinHaraTasaki15}) happens extremely fast. In the words of Goldstein \etal \cite{GoldsteinHaraTasaki15}, ``{\em what needs to be explained is, not that macroscopic systems approach
thermal equilibrium, but that they do so slowly.}'' The results of sections \ref{s:Ising} and \ref{s:LiebRobinson} suggest a possible direction to resolve this open problem: None of the proofs in \cite{Malabarba_etal14,GoldsteinHaraTasaki15,Reimann16} require or make use of a locality structure for the Hamiltonians or observables considered. Drawing randomly from all possible Hamiltonians or observables according to a ``natural'' probability distribution will most likely not describe models with interactions that act locally with finite-range or fast-decaying couplings. In this sense, the typical, randomly selected situations studied in \cite{Malabarba_etal14,GoldsteinHaraTasaki15,Reimann16} are expected to be dominated by long-range physics, and from the results reviewed in sections \ref{s:Ising} and \ref{s:LiebRobinson} it does not seem too surprising that the obtained typical timescales are extremely short. This line of thought also suggests a possible strategy for deriving physically realistic equilibration times, namely by imposing locality constraints on the classes of Hamiltonians and observables used in the typicality analysis. How such constraints can be implemented in practice is an open, and presumably difficult, question.

\section{Conclusions}

In this paper I have reviewed three seemingly disconnected topics, and provided context and discussions of the links between them. The first topic is the approach to equilibrium, where I have discussed a proof that generic isolated macroscopic quantum systems approach equilibrium in a probabilistic sense after a sufficiently long time. In the second part I have discussed an exact analytic solution for the relaxation dynamics of the quantum Ising model with long-range interactions, showing that, for long-range exponents $0\leqslant\alpha<1$, the timescale of equilibration scales like a power law with the system size $N$ and vanishes in the thermodynamic limit $N\to\infty$. Going beyond the study of a specific model system, a similar scaling behaviour was reported in the third part of this paper for general quantum spin models: Lieb-Robinson bounds for long-range systems with $0\leqslant\alpha<1$ make it appear plausible that the power law scaling of dynamical timescales with $N$ is a common occurrence in such systems. These observations suggest that locality (in contrast to the very long-ranged, nonlocal interactions of systems with small $\alpha$) is crucially affecting the equilibration timescales and needs to be considered when trying to derive physically realistic equilibration times by typicality techniques.

\ack
\addcontentsline{toc}{section}{Acknowledgements}
This work is based on, and reviews, work done jointly with my collaborators, in particular Romain Bachelard, Jens Eisert, Peter Reimann, David Storch, and Mauritz van den Worm, with all of whom I enjoyed pleasant and insightful discussions.
I acknowledge financial support by the National Research Foundation of South Africa through the Incentive Funding and the Competitive Programme for Rated Researchers.


\vspace{3mm}
\bibliographystyle{unsrt}
\bibliography{LRLR}

\end{document}